\documentclass[journal]{IEEEtran}

\usepackage[pdftex]{graphicx}
\usepackage[T1]{fontenc}
\usepackage{color}
\usepackage{varioref}
\usepackage{textcomp}
\usepackage{amsthm}
\usepackage{amsmath}
\usepackage{amssymb}
\usepackage{graphicx}
\usepackage{setspace}
\usepackage{esint}
\usepackage{enumerate}
\usepackage{algorithm}
\usepackage{algpseudocode}
\usepackage{pifont}

\usepackage{enumitem}
\usepackage{enumerate}
\usepackage{enumitem}
\usepackage{cite}
\makeatletter

\newcommand{\Rmnum}[1]{\expandafter\@slowromancap\romannumeral #1@}
\makeatother
 
\newtheorem{theorem}{Theorem}

\newtheorem{remark}{Remark}

\newtheorem{corollary}{Corollary}

\providecommand{\propositionname}{Proposition}

\ifCLASSINFOpdf

\else
  
\fi

\usepackage[T1]{fontenc}
\usepackage{etoolbox}

\makeatletter

\patchcmd{\maketitle}{\@fnsymbol}{\@alph}{}{}  
\makeatother

\title{Fundamental Limits of Coded Caching: Improved Delivery Rate-Cache Capacity Trade-off}
\author{
  Mohammad Mohammadi Amiri,~\IEEEmembership{Student Member,~IEEE} and\thanks{The authors are with Imperial College London, London SW7 2AZ, U.K. (e-mail: m.mohammadi-amiri15@imperial.ac.uk; d.gunduz@imperial.ac.uk).}
  \and
  Deniz~G\"und\"uz,~\IEEEmembership{Senior Member,~IEEE}
}
\date{}
\begin{document}

\maketitle

\begin{abstract}
A \textit{centralized coded caching} system, consisting of a server delivering $N$ popular files, each of size $F$ bits, to $K$ users through an error-free shared link, is considered. It is assumed that each user is equipped with a local cache memory with capacity $MF$ bits, and contents can be proactively cached into these caches over a low traffic period; however, without the knowledge of the user demands. During the peak traffic period each user requests a single file from the server. The goal is to minimize the number of bits delivered by the server over the shared link, known as the \textit{delivery rate}, over all user demand combinations. A novel coded caching scheme for the cache capacity of $M= (N-1)/K$ is proposed. It is shown that the proposed scheme achieves a smaller delivery rate than the existing coded caching schemes in the literature when $K > N \ge 3$. Furthermore, we argue that the delivery rate of the proposed scheme is within a constant multiplicative factor of $2$ of the optimal delivery rate for cache capacities $1/K \le M \le (N-1)/K$, when $K > N \ge 3$. 
\\ 
\end{abstract}

\begin{IEEEkeywords}
Centralized coded caching, network coding, proactive caching.
\end{IEEEkeywords}

\section{Introduction}
The\makeatletter{\renewcommand*{\@makefnmark}{}
\footnotetext{This work is funded partially by the European Research Council Starting Grant project BEACON (Project number 677854), and by the European Union's Horizon 2020 Research and Innovation Programme under grant agreement 690893, project TACTILENet: Towards Agile, effiCient, auTonomous and massIvely LargE Network of things.}\makeatother} growing number of users and their increasing appetite for high data rate content leads to network traffic congestion, particularly during peak traffic periods, whereas the resources are often underutilized during off-peak periods. Exploiting increasingly low-cost and abundant local storage capacity to proactively cache content at user devices is an effective way to utilize channel resources during off-peak hours, and mitigate the burden of high network load at times of heavy demand \cite{DowdyCaching,AlmerothCacing}. 

In order to model this dichotomy between peak and off-peak traffic periods, recent works on coded content caching consider two phases: In the \emph{placement phase}, which corresponds to periods of low network traffic, the cache memory of each user is filled by a central server without the knowledge of users' future demands. The main limitation of this phase is the capacity of users' caches. All user requests are revealed simultaneously during the peak traffic period. It is assumed that each user requests a single file from among a finite database of popular contents. Then the \emph{delivery phase} follows, in which a common message is transmitted to all the users in the system over an error-free shared channel. Each user tries to reconstruct the file it requests using its local cache content as well as the bits delivered by the server over the shared link. The goal of the server is to make sure that all the user demands, no matter which files are requested by the users, are satisfied at the end of the delivery phase. For a given number of files in the database, and given cache sizes at the users, the minimum rate, referred to as \textit{delivery rate}, at which data must be delivered through the shared link, independently of users' particular demands, is considered as the performance measure of a coded caching algorithm. Our goal is to design the placement and delivery phases jointly in order to minimize the delivery rate.    

In the case of uncoded caching, parts of popular contents are stored in the local cache memories, and once the user requests are revealed, remaining parts are delivered by the server over the shared link. The corresponding gain relative to not having a cache is called \emph{local caching gain}, and depends on the local cache capacity \cite{BorstCaching,GungorProactive}. On the other hand, it has been shown in \cite{MaddahAliCentralized} that \textit{coded caching} provides a \emph{global caching gain}, where users benefit not only from their own local cache, but also from the available cache memory across the network. Coded caching provides a novel method to mitigate network congestion during peak traffic hours by creating and exploiting coded multicasting opportunities across users. 

In a \emph{centralized coded caching} scheme, it is assumed that the central server knows the exact number of users in the system, and carefully places contents in the user caches during off-peak hours. A novel centralized coded caching scheme for a network of $K$ users requesting $N$ popular files of the same size is proposed in \cite{MaddahAliCentralized}, which has been shown to be optimal when the cache placement phase is uncoded and $K \ge N$ \cite{KaiWanUncodedOpt}. Authors in \cite{ZhiChenXOR} consider an alternative coded caching scheme, which was originally proposed in \cite{MaddahAliCentralized} for three users, and show that it is optimal when the number of users, $K$, is not less than the number of popular files, i.e., $N \le K$, and the normalized cache capacity $M$ satisfies $M \le 1/K$, i.e., a relatively small cache size. The delivery rate is further improved by the coded caching schemes investigated in \cite{KaiWanUncodedCaching} and \cite{MohammadDenizISITA}. Theoretical lower bounds on the delivery rate have also been derived to characterise the optimal performance of a caching system \cite{MaddahAliCentralized,SenguptaCaching,GhasemiCachingLowerBound,TianCaching}. In general, the minimum delivery rate for coded caching remains an open problem even in the symmetric setting considered in the aforementioned previous works.                     

The scheme of \cite{MaddahAliCentralized} has been extended to the decentralized setting in \cite{MaddahAliDecentralized}, in which the identity of the users requesting files during the delivery phase are not known in advance to perform the placement in coordination across caches. It has been further extended to multi-layer caching \cite{KaramchandaniHierarchical}, caching files with distinct sizes \cite{ZhangDistinctFileSizes} and popularities \cite{NiesenNonuniform, JiArXivNonuniform}, caching to users with distinct cache capacities \cite{WangHeterogenous,Asilomar16}, hierarchical coded caching \cite{KaramchandaniHierarchical,KonstantinosHierarchical}, caching files with lossy reconstruction \cite{QianDenizLossy, ElzaDistortionMemoryTradeoff, TimoDistortionCaching}, online cache placement \cite{PedarsaniOnlineCaching}, and delivery over a noisy shared link \cite{TimoErasureChannel,SaeediBidokhtiNoisy,HuangFadingChannelcodedcaching}. Similar caching techniques have also been employed in various other applications, e.g., device-to-device caching \cite{GregoryDtoD,JiCaireMolischDtoD} and femtocaching \cite{ShanmugamFemtoCaching,GolrezaeiFemtocachingDtoD}.

In this paper, we build upon our previous work in \cite{MohammadDenizISITA}, and propose a novel centralized coded caching scheme, when the cache capacity of the users is given by $M=(N-1)/K$. This new caching scheme utilizes coded content placement, in which contents are partitioned into smaller chunks, and pairwise XOR-ed contents are placed in the user caches. The delivery phase utilizes both coded and uncoded transmission. We show that the proposed caching scheme requires a smaller delivery rate (evaluated for the worst-case user demands) compared to the best achievable scheme in the literature for the same cache capacity, when $N<K$. We then extend the improvement in the delivery rate to a larger range of cache capacities utilizing the memory-sharing argument. Finally, we show that the delivery rate achieved by the proposed caching scheme is within a constant multiplicative factor of $2$ of the optimal delivery rate for cache capacities satisfying $1/K \le M \le (N-1)/K$, when $K > N \ge 3$. We believe that the ideas behind our centralized coded caching scheme may lead to further improvements on the delivery rate in decentralized as well as online caching systems. 

We remark that the proposed caching scheme improves the performance upon the state-of-the-art when there are more users in the system than the number of files in the database, i.e., $N<K$, and cache capacity of each user is relatively small. This scenario is valid for contents that become highly popular over the Internet, and are demanded by a huge number of users, each equipped with a cache memory of comparatively small size, in a relatively short time interval, for example, viral videos distributed over social networks, new episodes of popular TV series, breaking news videos, or for broadcasting different software updates to millions of clients.    

The rest of this paper is organized as follows. In Section \ref{SysModel}, we investigate the system model, and present the relevant previous results in the literature. In Section \ref{SecondScheme}, a novel centralized coded caching scheme is proposed, and its delivery rate is analyzed and compared with the state-of-the-art results both theoretically and numerically. All the proofs can be found in the Appendix. Finally, conclusions are included in Section \ref{Conc}.            

\section{System Model and Previous Results}\label{SysModel}

\begin{figure}[!t]
\centering
\includegraphics[scale=0.68]{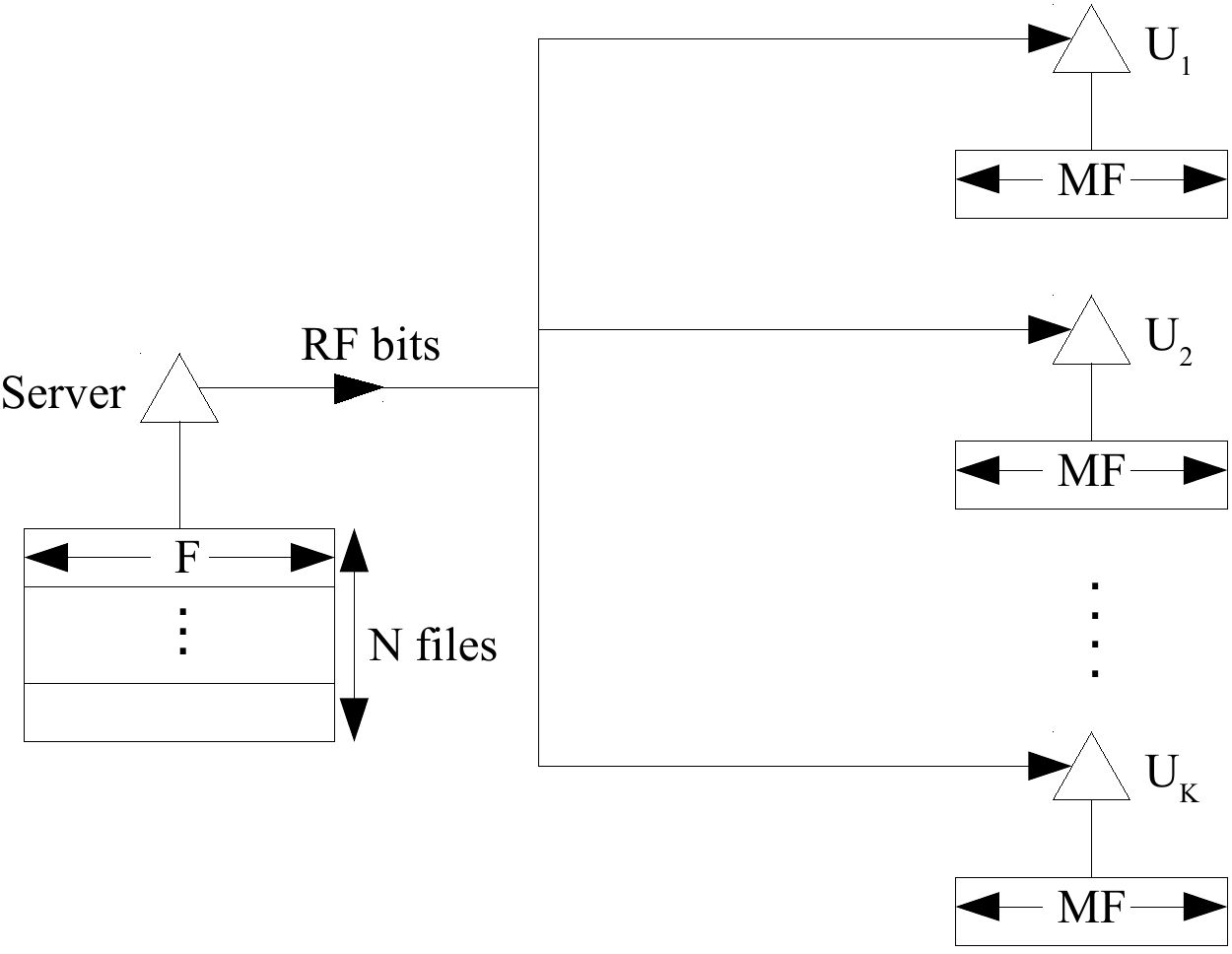}
\caption{Illustration of a centralized caching system consisting of a server with a database of $N$ popular files, each with size $F$ bits, serving $K$ users, each with a cache of capacity $MF$ bits, requesting a single file from the database. These requests are served simultaneously through an error-free shared link.} 
\label{System_Model}
\end{figure}

We assume that there is a central server broadcasting data to $K \in \mathcal{Z}^{+}$ users, $U_{1}, ..., U_{K}$, through a shared error-free link, where $\mathcal{Z}^{+}$ is the set of positive integers. As depicted in Fig. \ref{System_Model}, the server has $N \in \mathcal{Z}^{+}$ files in its database, each with the size of $F$ bits, that is, file $n$, denoted by $W_n$, for $n=1,...,N$, is a random variable uniformly distributed over $\left[ {{2^F}} \right] \buildrel \Delta \over = \left\{ 1,...,2^F \right\}$. We denote the whole database by $\mathbf{W} \buildrel \Delta \over = \left( {{W_1},...,{W_N}} \right)$. Each user is assumed to have a local cache of capacity $MF$ bits.

Similarly to \cite{MaddahAliCentralized}, the system operates in two phases. In the \textit{placement phase}, each user's cache is filled with bits depending on the whole database and the capacity of the user caches, $M$. The content of user $k$'s cache at the end of the placement phase is denoted by $Z_k$. At the end of the placement phase, each user requests one of the files from the database. We use $\mathbf{d} = \left( {{d_1},...,{d_K}} \right)$ to denote the demand vector, where $d_k \in \left[ N \right]$ corresponds to the demand of user $U_k$. In the \textit{delivery phase}, the server, having received the requests of all the users, transmits a common message of size $RF$ bits over the shared link. Note that, in the centralized caching model considered here, this common message depends not only on the contents in the database and the user requests, but also on the contents of users' caches. At the end of the delivery phase, each user $k$ tries to decode its requested file $W_{d_k}$ using its cache content $Z_k$, and the common message received over the shared link.

\textbf{Definition.} An $\left( M,R,F \right)$ \textit{caching and delivery code} for the above caching system with $K$ users and $N$ files consists of 
\begin{enumerate}[label=\roman*)]
\item $K$ caching functions:
\begin{equation}\label{S1} {\phi _k}:{\left[ {{2^F}} \right]^N} \to \left[ {{2^{\left\lfloor {FM} \right\rfloor }}} \right],
\end{equation}
which maps the database $\mathbf{W}$ to the cache content $Z_k$ of user $k$, i.e., $Z_k={\phi _k}\left( \mathbf{W} \right)$;
\item delivery encoding function:
\begin{equation}\label{S2} \psi :{\left[ {{2^F}} \right]^N} \times {\left[ N \right]^K} \to \left[ {{2^{\left\lfloor {FR} \right\rfloor }}} \right],
\end{equation}
which maps the database $\mathbf{W}$ and the particular demand vector $\mathbf{d}$ to a message $X$ over the shared link, i.e., $X=\psi \left( \mathbf{W}, \mathbf{d} \right)$; 
\item $K$ decoding functions:
\begin{equation}\label{S3} {\mu _k}:\left[ {{2^{\left\lfloor {FM} \right\rfloor }}} \right] \times \left[ {{2^{\left\lfloor {FR} \right\rfloor }}} \right] \times \left[ N \right]^K \to \left[ {{2^F}} \right],
\end{equation}
which maps the cache content $Z_k$, message over the shared link $X$, and the demand vector $\mathbf{d}$ to the reconstructed message ${{\hat W}_k}$ at user $k$, i.e., we have ${{\hat W}_k}={\mu _k}\left(Z_k, X, \mathbf{d} \right)$. 
\end{enumerate}

The probability of error of an $\left( M,R,F \right)$ caching and delivery code is defined as 
\begin{equation}\label{S4} {P_e} \buildrel \Delta \over = \mathop {\max }\limits_{\left( {{d_1},{d_2},...,{d_K}} \right)} \Pr \left\{ {\mathop  \bigcup_{k = 1}^K \left\{ {{{\hat W}_k} \ne {W_{{d_k}}}} \right\}} \right\} \cdot
\end{equation}

In this model, $M$ is the normalized cache capacity while $R$ is the delivery rate, which corresponds to the number of bits transmitted over the shared link, also normalized by the file length $F$.

\textbf{Definition.} The delivery rate-cache capacity pair $\left( R,M \right)$ is \textit{achievable} if for any $\varepsilon  > 0$, there exist a large enough $F$ and a corresponding $\left( M,R,F \right)$ caching and delivery code with ${P_e} < \varepsilon $.

There is a trade-off between the cache capacity of the users and the delivery rate. For example, when $M = 0$, in the worst case, users request as distinct files as possible, and the server has to transmit all the requests over the shared link, that is, the delivery rate has to be at least $R = \min \left\{ {N,K} \right\}$. In the other extreme case, when the cache capacities are large enough to store all the $N$ files, i.e., when $M = N$, all the requests can be satisfied directly from the local caches, and the delivery rate can be zero, i.e., $R=0$. In general, we define the delivery rate-cache capacity trade-off $R^*(M)$ as follows: 
\begin{equation}\label{S5} {R^*}\left( M \right) \buildrel \Delta \over = \inf \left\{ {R:\left( {R,M} \right) \mbox{ is achievable} } \right\} \cdot
\end{equation}
Our goal is to obtain the delivery rate-cache capacity trade-off for all possible cache capacity values in between the above extreme scenarios by designing the placement and delivery phases jointly. 

When $N > K$, the centralized coded caching scheme proposed by Maddah-Ali and Niesen in \cite{MaddahAliCentralized} for cache capacities $M = tN/K$, for $t = 1, ..., K$, is the best known achievable scheme in the literature. On the other hand, when $N \le K$, the scheme proposed in \cite{ZhiChenXOR} is shown to be optimal for $M \le 1/K$. In this case, to characterize the best achievable delivery rate in the literature for $M \ge 1/K$, we calculate the delivery rate achieved by memory-sharing between the schemes proposed for $M=1/K$ in \cite{ZhiChenXOR} and for $M = tN/K$ in \cite{MaddahAliCentralized}, for $t \in \left[ K \right]$, as follows:
\begin{equation}\label{MemorySharingDelRate}
{R_{M,t}} = {\alpha_t}M + {\beta_t}, \quad \mbox{for $\frac{1}{K} \le M \le \frac{{tN}}{K}$},
\end{equation}
where 
\begin{subequations}
\label{MemorySharingAlphaBeta}
\begin{align} \label{Alpha}
{\alpha_t} &= \frac{{K\left( {K - t} \right)}}{{\left( {t + 1} \right)\left( {tN - 1} \right)}} - \frac{{N\left( {K - 1} \right)}}{{tN - 1}},\\
{\beta_t} &= N - \frac{N}{K} - \frac{{K - t}}{{\left( {t + 1} \right)\left( {tN - 1} \right)}} + \frac{{N\left( {K - 1} \right)}}{{K\left( {tN - 1} \right)}}.
\label{Beta}
\end{align}
\end{subequations}
The value of $t \in \left[ K \right]$ that minimizes $\alpha_t$ in \eqref{Alpha} is denoted by $t^*$: 
\begin{equation}\label{TStar}
t^* \buildrel \Delta \over = \arg \mathop {\min }\limits_{t \in \left[ K \right]} \left\{ {\alpha_t} \right\}.
\end{equation}
This leads to the straight line $R_{M,t}$ in \eqref{MemorySharingDelRate} with the lowest slope, which characterizes a range of delivery rates achieved through memory-sharing between the schemes proposed for $M=1/K$ in \cite{ZhiChenXOR} and for $M=t^*N/K$ in \cite{MaddahAliCentralized}. This scheme, referred to as the Maddah-Ali-Niesen-Chen (MNC) scheme, is considered as the state-of-the-art for $1/K \le M \le t^*N/K$ throughout this paper, and achieves a delivery rate of
\begin{equation}\label{RbTStar}
{R_b}\left( M \right) = R_{M,t^*} = {\alpha_{t^*}}M + \beta_{t^*}, \quad \mbox{for $\frac{1}{K} \le M \le \frac{{t^*N}}{K}$}.
\end{equation}
For $M \ge t^*N/K$, Maddah-Ali and Niesen's scheme in \cite{MaddahAliCentralized} again achieves the best performance in the literature.

In the following, we will introduce the placement and delivery phases for the proposed caching strategy for a cache capacity of $M = (N-1)/K$. We will show that, for this particular cache capacity the proposed scheme achieves a smaller delivery rate than the state-of-the-art presented above. We then characterize a delivery rate-cache capacity trade-off using memory-sharing between our scheme and the existing ones, which extends the improvement to a larger range of cache capacity values.   

\section{Proposed Coded Caching Scheme}\label{SecondScheme}

Before we present a detailed description and analysis of the proposed coded caching scheme, we illustrate it on a simple example highlighting its main ingredients. This example will allow us not only to provide the intuition behind the proposed caching scheme, but also to show its superiority over the MNC scheme.

\theoremstyle{definition}
\newtheorem{exmp}{Example}

\begin{exmp}\label{exampleGeneralCase}
Consider a caching system with a database of $N=3$ files, $W_1$, $W_2$ and $W_3$. There are $K=5$ users in the system, each of which is equipped with a cache of capacity $M = (N-1)/K = 2/5$. To perform the placement phase, each file $W_i$, $\forall i$, is first divided into $K=5$ non-overlapping subfiles $W_{i,j}$, each of the same length $F/5$ bits, for $j = 1, ..., 5$. The following contents are then cached by user $U_k$, $\forall k$, in the placement phase:
\begin{equation}\label{Example3CachePlacement}
{Z_k} =  \left( W_{1,k} \oplus W_{2,k}, W_{2,k} \oplus W_{3,k} \right),
\end{equation}
where $\oplus$ is the bitwise XOR operation. Since each subfile $W_{i,j}$ has a length of $F/5$ bits, the cache placement phase satisfies the memory constraint. See Fig. \ref{CacheContent} for an explicit illustration of the cache contents at the end of the placement phase.   

\begin{figure}[!t]
\centering
\includegraphics[scale=0.65]{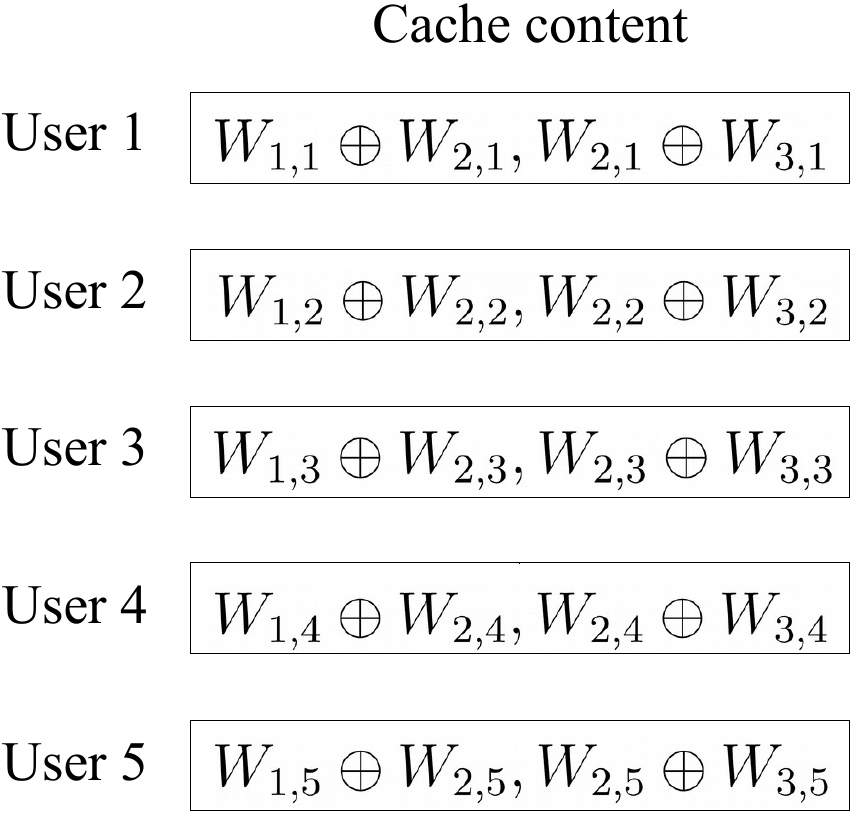}
\caption{Cache placement for the proposed coded caching scheme for $N=3$ files and $K=5$ users, each equipped with a cache of capacity $M=2/5$.} 
\label{CacheContent}
\end{figure} 

We argue that for the proposed caching scheme with $N<K$, the worst-case user demands happens when each file is requested by at least one user. This fact will later be clarified in Remark \ref{RemWorstCaseDemand}. By re-labeling the files and re-ordering the users, without loss of generality, the user demand combination is assumed to be $\mathbf{d}=\left( 1,1,1,2,3 \right)$.

In the delivery phase, each subfile $W_{i,j}$, $\forall i,j$, is further divided into $N-1=2$ distinct pieces $W_{i,j}^{\left( l \right)}$, for $l=1, 2$, each of size $F/10$ bits, i.e., ${W_{i,j}} = \left( {W_{i,j}^{\left( 1 \right)},W_{i,j}^{\left( 2 \right)}} \right)$, $\forall i, j$. Accordingly, cache contents \eqref{Example3CachePlacement} can be rewritten as 
\begin{equation}\label{Example3CachePlacementRewritten}
{Z_k} = \bigcup\limits_{l = 1}^2 {\left( W_{1,k}^{(l)} \oplus W_{2,k}^{(l)}, W_{2,k}^{(l)} \oplus W_{3,k}^{(l)} \right)}.
\end{equation}
The contents are then delivered by the server in three different parts. The following contents are sent in each part of the delivery phase:
\begin{enumerate}[label=\bfseries Part \arabic*:,align=left]
\item $W_{2,1}^{\left( 1 \right)}$, $W_{3,1}^{\left( 2 \right)}$, $W_{2,2}^{\left( 1 \right)}$, $W_{3,2}^{\left( 2 \right)}$, $W_{2,3}^{\left( 1 \right)}$, $W_{3,3}^{\left( 2 \right)}$, $W_{1,4}^{\left( 1 \right)}$, $W_{3,4}^{\left( 2 \right)}$, $W_{1,5}^{\left( 1 \right)}$, $W_{2,5}^{\left( 2 \right)}$,
\item ${W_{1,1}} \oplus {W_{1,2}}$, ${W_{1,2}} \oplus {W_{1,3}}$,
\item $W_{2,1}^{\left( 2 \right)} \oplus W_{2,2}^{\left( 2 \right)}$, $W_{2,2}^{\left( 2 \right)} \oplus W_{2,3}^{\left( 2 \right)}$, $W_{1,4}^{\left( 2 \right)} \oplus W_{2,3}^{\left( 2 \right)}$, $W_{3,1}^{\left( 1 \right)} \oplus W_{3,2}^{\left( 1 \right)}$, $W_{3,2}^{\left( 1 \right)} \oplus W_{3,3}^{\left( 1 \right)}$, $W_{1,5}^{\left( 2 \right)} \oplus W_{3,3}^{\left( 1 \right)}$, $W_{2,5}^{\left( 1 \right)} \oplus W_{3,4}^{\left( 1 \right)}$.  
\end{enumerate}

Having received the contents delivered in part 1, each user can retrieve all the subfiles placed in its own cache in XOR-ed form. For example, user $U_1$ can decode all the subfiles $W_{i,1}$, for $i=1, 2, 3$, after receiving the pair $\left( {W_{2,1}^{\left( 1 \right)},W_{3,1}^{\left( 2 \right)}} \right)$. 

With the second part, each user can obtain the subfiles of its desired file that have been cached by another user with the same demand. For example, the contents ${W_{1,1}} \oplus {W_{1,2}}$ and ${W_{1,2}} \oplus {W_{1,3}}$ help users $U_1$, $U_2$ and $U_3$ to obtain the subfiles of their request, $W_1$, which have been cached by each other. 

Finally, the last part of the delivery phase enables each user to decode the missing pieces of its requested file having been cached by another user with a different demand. For example, the delivered contents $W_{2,1}^{\left( 2 \right)} \oplus W_{2,2}^{\left( 2 \right)}$, $W_{2,2}^{\left( 2 \right)} \oplus W_{2,3}^{\left( 2 \right)}$, and $W_{1,4}^{\left( 2 \right)} \oplus W_{2,3}^{\left( 2 \right)}$ help users $U_1$, $U_2$, and $U_3$ to obtain the piece $W_{1,4}^{\left( 2 \right)}$, and user $U_4$ can also decode the pieces $W_{2,1}^{\left( 2 \right)}$, $W_{2,2}^{\left( 2 \right)}$, and $W_{2,3}^{\left( 2 \right)}$. It can be verified that having received all the bits sent in three parts in the delivery phase, each user can obtain its desired file with a total delivery rate of $R_c=2.1$. On the other hand, the MNC scheme achieves a delivery rate of $R_b=2.12$ for the setting under consideration.    
\qed
\end{exmp}

In the sequel, we present the cache placement and delivery phases of the proposed scheme in the general setting, analyze its delivery rate, and compare it with the state-of-the-art. We will observe that its superiority over the MNC scheme is not limited to the particular setting in the above example. 

\subsection{Placement Phase}\label{SecondProposedSchemePlacement}

We first generate $K$ non-overlapping subfiles, each of size $F/K$ bits, for each file $W_i$, $\forall i$, denoted by $W_{i,1}, \ldots, W_{i,K}$. Similarly to \cite{ZhiChenXOR} we use coded placement; that is, contents are cached in XOR-ed form in the placement phase. However, unlike in \cite{ZhiChenXOR}, instead of XORing subfiles of all the files in the database, we XOR subfiles in pairs. In particular, the following contents are cached by user $U_k$, for $k=1, ..., K$, in the placement phase:  
\begin{equation}\label{SecondSchemeCachePlacement}
{Z_k} = \bigcup\limits_{i = 1}^{N - 1} {\left( {{W_{i,k}} \oplus {W_{i + 1,k}}} \right)} .
\end{equation}
Since each subfile has a size of $F/K$ bits, the limited memory of each cache is filled completely by the proposed placement scheme. In this way, each subfile of all the files is cached by exactly one user in the XOR-ed form. Hence, the whole of each file can be found in the caches of the users across the network (in coded form).  

\subsection{Delivery Phase}\label{SecondProposedSchemeDeilvery}

Note that, in the proposed caching scheme all the database is stored across the caches of the users. Therefore, in the delivery phase, the server first transmits the appropriate subfiles so that each user can recover all the subfiles stored in its cache in XOR-ed form. Then, the server transmits XOR of contents that are available at two different users, where each content is requested by the other user. This, equivalently, enables the two users to exchange their contents. By appropriately pairing subfiles, the server guarantees that each user receives the subfiles of its requested file that have been cached by every other user in the system.   

Without loss of generality, by re-ordering the users, it is assumed that the first $K_1$ users, referred to as the group $G_1$, request file $W_1$, the next $K_2$ users, which form the group $G_2$, demand $W_2$, and so on so forth. For notational convenience, we define  
\begin{equation}\label{SiDefinition}
{S_i} \buildrel \Delta \over = \sum\limits_{l = 1}^i {{K_l}},
\end{equation}
where we set $S_0 \buildrel \Delta \over=0$. Thus, the general-case user demands can be expressed as follows:
\begin{equation}\label{UserDemandsGeneralCase}
d_k=i, \quad S_{i-1}+1 \le k \le S_{i},  \mbox{ for $i=1, ..., N$}.
\end{equation} 
It is illustrated in Appendix \ref{ProofSecondTheorem} that the delivery rate of the proposed coded caching scheme does not depend on $K_i$, $\forall i$, i.e., the proposed scheme is not affected by the popularity of the files, as long as $K_i > 0$. Therefore, when $N<K$, the worst-case user demands for the proposed scheme happens when each file is requested by at least one user, i.e., $K_i \ge 1$, for $i=1, ..., N$.

The proposed delivery phase is divided into three distinct parts, and the contents delivered in part $i$ is denoted by $X_i$, for $i=1,2,3$. Hence, $X = \left( X_1, X_2, X_3 \right)$ is transmitted over the shared link in the delivery phase. The delivery phase algorithm is presented for the worst-case user demands when $N<K$, i.e., when there is at least one user requesting each file. The proposed delivery phase algorithm is then extended to all values of $N$ and $K$ for a generic user demand combination assumption by introducing a new variable $N'$ as the total number of files requested by the users.

To symmetrize the contents transmitted in the delivery phase, this phase is performed by further partitioning each subfile; that is, each subfile $W_{i,j}$, $\forall i,j$, is divided into $(N-1)$ distinct pieces $W_{i,j}^{\left( 1 \right)}, \ldots, W_{i,j}^{\left( N-1 \right)}$, each of length $F/(K(N-1))$ bits. Considering these smaller pieces, the content placed in the cache of user $U_k$, for $k=1, ..., K$, can be re-written as follows:
\begin{equation}\label{SecondSchemeCachePlacementRewrite}
{Z_k} = {\bigcup\limits_{l = 1}^{N - 1} {\bigcup\limits_{i = 1}^{N - 1} {\left( {W_{i,k}^{\left( l \right)} \oplus W_{i + 1,k}^{\left( l \right)}} \right)} } } .
\end{equation}

\begin{algorithm}
\caption{Part 1 of the delivery phase}
\label{Part1cgeneralSecond}
\begin{algorithmic}[1]
\Statex
\Procedure {Per-user coding}{}
\For{$i = 1, \ldots, N$ }
\For{$k = S_{i-1}+1, \ldots, S_i$ }
\For{$j = 1, \ldots, N$ and $j \ne i$}
\State $m_{j,k} =
\begin{cases} 
{j,\quad \quad \;\;\; j < i}\\
{j-1,\quad j>i}
\end{cases}$
\State $X_1 \leftarrow \left(X_1,W_{j,k}^{\left(m_{j,k}\right)}\right)$
\EndFor
\EndFor
\EndFor
\EndProcedure
\end{algorithmic}
\end{algorithm}

The first part of the delivery phase is stated in Algorithm \ref{Part1cgeneralSecond}. The main purpose of this part is to enable each user $U_k$, $\forall k$, to retrieve all the subfiles $W_{j,k}$, for $j=1, ..., N$, that have been cached in the cache of user $U_k$ in XOR-ed form. We remark that according to the cache placement in \eqref{SecondSchemeCachePlacementRewrite}, for each $l \in \left[ N-1 \right]$, by delivering only one of the pieces $W_{1,k}^{\left( l \right)},...,W_{N,k}^{\left( l \right)}$, user $U_k$ can recover all the pieces $W_{1,k}^{\left( l \right)},...,W_{N,k}^{\left( l \right)}$, $\forall k$. Hence, each user requires a total of $(N-1)$ distinct pieces to recover all the subfiles placed in its cache in XOR-ed form. To perform an efficient and symmetric delivery phase, $(N-1)$ distinct pieces, which are in the cache of user $U_k$ in XOR-ed form, corresponding to $(N-1)$ different subfiles of the files that are not requested by user $U_k$ are delivered to that user. For example, for user $U_1$ requesting file $W_1$, the pieces $\left( {W_{2,1}^{\left( 1 \right)},W_{3,1}^{\left( 2 \right)},...,W_{N,1}^{\left( {N - 1} \right)}} \right)$ are delivered by Algorithm \ref{Part1cgeneralSecond}. Accordingly, for user $U_k$ that has requested file $W_i$, the pieces $\left( {W_{1,k}^{\left( 1 \right)},...,W_{i - 1,k}^{\left( {i - 1} \right)},W_{i + 1,k}^{\left( i \right)},...,W_{N,k}^{\left( {N - 1} \right)}} \right)$ are delivered over the shared link. Thus, each user $U_k$ can recover all subfiles $W_{j,k}$, $\forall j \in \left[ N \right]$, stored in its cache in XOR-ed form. Note also that, Algorithm \ref{Part1cgeneralSecond} delivers at most one piece of each subfile over the shared link. In Algorithm \ref{Part1cgeneralSecond}, we denote the index of the piece of subfile $W_{j,k}$, that is delivered in part 1 of the delivery phase by $m_{j,k}$. We will later refer to these indexes in explaining the other parts of the delivery phase. Note that, the pieces $\left( {W_{1,k}^{\left( 1 \right)},...,W_{i - 1,k}^{\left( {i - 1} \right)},W_{i + 1,k}^{\left( i \right)},...,W_{N,k}^{\left( {N - 1} \right)}} \right)$ are targeted for user $U_k$ in group $G_i$ demanding file $W_i$, for $i=1, ..., N$, and $k=S_{i-1}+1, ..., S_i$. Accordingly, for $j \in \left[ N \right]\backslash \left\{ i \right\}$, we have 
\begin{equation}\label{PieceIndex}
m_{j,k} =
\begin{cases} 
{j,\quad \quad \;\;\; j < i},\\
{j-1,\quad j>i,}
\end{cases}
\end{equation}
which results in $m_{j,k} \le N-1$.

For example, in Example \ref{exampleGeneralCase} given above, we have $\left( {m_{1,4}},{m_{1,5}} \right) = \left( {1,1} \right)$, $\left( {m_{2,1}},{m_{2,2}},{m_{2,3}}, {m_{2,5}} \right) = \left( {1,1,1,2} \right)$, and $\left( {{m_{3,1}},{m_{3,2}},{m_{3,3}},{m_{3,4}}} \right) = \left( {2,2,2,2} \right)$. 

Algorithm \ref{Part2cgeneralSecond} presents the second part of the proposed delivery phase, which allows each user to obtain its missing subfiles that have been cached by the other users in the same group. Note that, having received part 1 of the delivery phase, user $U_j$ in group $G_i$, for $i=1, ..., N$, and $j = S_{i-1}+1, ..., S_i$, can recover subfile $W_{i,j}$. Algorithm \ref{Part2cgeneralSecond} delivers $\bigcup \limits_{k = S_{i - 1} + 1}^{S_{i}-1} \left( W_{i,k} \right.$ $\oplus$ $\left. W_{i,k + 1} \right)$, with which user $U_j$ can recover all the subfiles ${W_{i,{S_{i - 1}} + 1}}, \ldots, {W_{i,{S_{i}}}}$, i.e., the subfiles of file $W_i$ placed in the caches of users in group $G_i$.

\begin{algorithm}
\caption{Part 2 of the delivery phase}
\label{Part2cgeneralSecond}
\begin{algorithmic}[1]
\Statex
\Procedure {Inter-group coding}{}
\For {$i = 1, \ldots, N$}
\State {$X_2 \leftarrow \left( X_2, \mathop  \bigcup \limits_{k = S_{i - 1} + 1}^{S_{i}-1} \left( {{W_{i,k}} \oplus {W_{i,k + 1}}} \right) \right)$}
\EndFor
\EndProcedure
\end{algorithmic}
\end{algorithm}

The last part of the proposed delivery phase is presented in Algorithm \ref{Part3cgeneralSecond}, with which each user can receive the missing pieces of its desired file that have been placed in the cache of users in other groups. We deliver these pieces by exchanging data between the users in different groups. Observe that, for each user in group $G_i$, for $i=1, ..., N$, one piece of the subfile of its requested file $W_i$ which is available to the users in $G_j$, for $j=1,...,N, j \ne i$, was delivered in the first part of the delivery phase. Therefore, there are $(N-2)$ missing pieces of a file requested by a user, which have been placed in the cache of a user in a different group. For example, by delivering the pieces $\left( {W_{2,1}^{\left( 1 \right)},W_{3,1}^{\left( 2 \right)},...,W_{N,1}^{\left( {N - 1} \right)}} \right)$ to user $U_1$ demanding file $W_1$ in part 1 of the delivery phase, each user in group $G_j$ with demand $W_j$ can obtain the piece ${W_{j,1}^{\left( {j - 1} \right)}}$, for $j=2, ..., N$. Therefore, there are $(N-2)$ missing pieces of the files requested by the users in groups $G_2, ..., G_N$, that are available in the cache of user $U_1$. Consider exchanging data between each user $U_p$ in group $G_i$ (demanding file $W_i$) and each user $U_q$ in group $G_j$ (demanding file $W_j$), for $i=1, ..., N-1$ and $j=i+1, ..., N$, where $p=S_{i-1}+1, ..., S_i$ and $q=S_{j-1}+1, ..., S_j$. The subfile cached by user $U_p$ ($U_q$) requested by user $U_q$ ($U_p$) is $W_{j,p}$ ($W_{i,q}$). According to \eqref{PieceIndex}, the index of the piece of subfile $W_{j,p}$ ($W_{i,q}$) delivered in the first part of the delivery phase is equal to $m_{j,S_{i}}$ ($m_{i,S_{j}}$), $\forall p \in \left\{ {{S_{i - 1}} + 1,...,{S_i}} \right\}$ and $\forall q \in \left\{ {{S_{j - 1}} + 1,...,{S_j}} \right\}$. Hence, the indexes of the missing pieces of each user in group $G_i$ ($G_j$) available in the cache of a user in group $G_j$ ($G_i$) are $\left[ {N - 1} \right] \backslash \left\{ m_{i,S_j} \right\}$ $\left( \left[ {N - 1} \right] \backslash \left\{ m_{j,S_i} \right\} \right)$. Let $\pi _1^{i,j} (\cdot)$ and $\pi _2^{i,j} (\cdot)$ be arbitrary permutations on sets $\left[ {N - 1} \right] \backslash \left\{ m_{i,S_j} \right\}$ and $\left[ {N - 1} \right] \backslash \left\{ m_{j,S_i} \right\}$, respectively, for $i=1, ..., N-1$ and $j=i+1, ..., N$. For $m_1 = \pi _1^{i,j} (l)$ and $m_2 = \pi _2^{i,j} (l)$, $\forall l \in \left[ N-2 \right]$, after receiving the corresponding contents delivered by Algorithm \ref{Part3cgeneralSecond}, all the users in $G_i$ can recover the pieces ${W_{i,{S_{j - 1}} + 1}^{(m_1)}}, \ldots, {W_{i,{S_{j}}}^{(m_1)}}$, and all the users in $G_j$ can recover the pieces ${W_{j,{S_{i - 1}} + 1}^{(m_2)}}, \ldots, {W_{j,{S_{i}}}^{(m_2)}}$.

\begin{algorithm}
\caption{Part 3 of the delivery phase}
\label{Part3cgeneralSecond}
\begin{algorithmic}[1]
\Statex
\Procedure {Intra-group coding}{}
\For {$i = 1, \ldots, N-1$}
\For {$j = i+1, \ldots, N$}
\For{$l = 1, \ldots, N-2$}
\State ${m_1} = \pi _1^{i,j} (l)$
\State ${m_2} = \pi _2^{i,j} (l)$
\State {$X_3 \leftarrow \left( X_3, \mathop  \bigcup \limits_{k = S_{j - 1} + 1}^{S_j - 1} \left( {{W_{i,k}^{\left( m_1 \right)}} \oplus {W_{i,k + 1}^{\left( m_1 \right)}}} \right), \qquad \qquad \qquad \qquad \qquad \qquad \qquad \qquad \right.$ 
$\left. \qquad \quad \; \mathop  \bigcup \limits_{k = S_{i - 1} + 1}^{S_i - 1} \left( {W_{j,k}^{\left( m_2 \right)}\oplus W_{j,k+1}^{\left( m_2 \right)}} \right), W_{i, S_j}^{\left( m_1 \right)} \oplus W_{j, S_i}^{\left( m_2 \right)} \right)$}
\EndFor
\EndFor
\EndFor
\EndProcedure
\end{algorithmic}
\end{algorithm}

Having received all three parts of the delivery phase, each user $U_k$, $\forall k$, can recover all the pieces of its desired file $W_{d_k}$ that have been placed in any of the caches in the system. Together with the proposed placement phase, which guarantees that all the subfiles of each file is available in one of the caches across the network, we can conclude that the demand of each user is satisfied by the proposed caching algorithm. It is to be noted that when $N=2$, the proposed scheme is equivalent to the one proposed in \cite{ZhiChenXOR}, so we consider $N \ge 3$ throughout this paper.  

\subsection{Delivery Rate Analysis}\label{DelRateAnalysisSecScheme}
The delivery rate of the proposed coded caching scheme is provided in the following theorem, whose detailed proof can be found in Appendix \ref{ProofSecondTheorem}.  

\begin{theorem}\label{TheoremSecondSchemeDelRate}  
In centralized caching with $N$ files, each of length $F$ bits, and $K$ users, each equipped with a cache of capacity $MF$ bits, if $N<K$ and $M=(N-1)/K$, the following worst-case delivery rate is achievable: 
\begin{equation} \label{TheoremSecondSchemeDeliveryRate}
{R_{{c}}}\left( {\frac{{N - 1}}{K}} \right) = N\left( {1 - \frac{N}{{2K}}} \right).
\end{equation}
\end{theorem}

\begin{remark}\label{RemWorstCaseDemand}
To perform the proposed delivery phase for the general case, without loss of generality, the user demand combination is assumed as in \eqref{UserDemandsGeneralCase}, such that $K_i \ge 1$, for $i \le N'$; and $K_i=0$, otherwise, for some $N' \le N$, that is a total of $N'$ files are requested by the users in the system. In this case, each subfile is divided into $(N'-1)$ distinct equal-length pieces, and in the delivery phase algorithm, the value $N$ is substituted by $N'$. Hence, according to the delivery rate analysis provided in Appendix \ref{ProofSecondTheorem}, all the users' demands can be satisfied by delivering a total number of $R_c=N'\left( {1 - N'/\left( {2K} \right)} \right)$ file(s). Since $R_c$ is an increasing function of $N'$, we can conclude that, for $N<K$, the worst-case user demands happens when all the $N$ files in the database are requested by at least one user, i.e., $K_i \ge 1$, for $i=1, ..., N$.
\end{remark}

\begin{remark}\label{NlargerthanequalK}
Based on Remark \ref{RemWorstCaseDemand}, when $N \ge K$, the worst-case user demands corresponds to the case when $N'=K$, i.e., all the users request distinct files in the database. Hence, the proposed scheme achieves a delivery rate of $K/2$ when $M = (N-1)/K$, which is equal to the delivery rate of the state-of-the-art for the same cache capacity when $N=K$ and $N=K+1$. However, the scheme proposed in \cite{MaddahAliCentralized} achieves a delivery rate smaller than $K/2$ for $M=(N-1)/K$, when $N \ge K+2$.  
\end{remark}

\begin{remark}\label{RemImprovedSecondScheme}
It is possible to show that the proposed scheme improves upon the MNC scheme for $M = (N-1)/K$. It is proven in Appendix \ref{ProofImprovementSecondScheme} that the delivery rate achieved by memory-sharing between the scheme presented in \cite{ZhiChenXOR} for $M=1/K$ and the scheme proposed here for $M=(N-1)/K$ has a smaller slope compared to the delivery rate of the MNC scheme at $M=1/K$, when $K > N \ge 3$. Since the MNC scheme is achieved through memory-sharing between the schemes proposed in \cite{ZhiChenXOR} and \cite{MaddahAliCentralized} for $M = 1/K < (N-1)/K$ and $M = t^*N/K > (N-1)/K$, respectively, where $t^* \ge 1$ is determined by \eqref{TStar}, it is concluded that $R_c \left( (N-1)/K \right) \le R_b \left( (N-1)/K \right)$, when $K > N \ge 3$. 
\end{remark}

\begin{remark}\label{IntuitiveImprovement}
We remark that the gain of the proposed scheme compared to the MNC scheme is due to the better use of the available cache capacities of the users when $MK=N-1$, and partitioning each subfile into a number of distinct pieces, which allows performing an efficient and symmetric delivery phase. The proposed coded cache placement allows each user to retrieve all the subfiles in its cache at a relatively small cost. Furthermore, each user receives the bits of all the requested files rather than receiving bits only of its requested file, creating symmetry across users, which is helpful for the later steps of the delivery phase, and leads to a reduction in the delivery rate.    
\end{remark}

Since $1/K < (N - 1)/K < {t^*}N/K$, the improvement of the proposed coded caching scheme for $M = (N-1)/K$ can be extended to the range of cache capacities $M \in \left( {1/K,{t^*}N/K} \right)$ through memory-sharing with the caching schemes proposed in \cite{ZhiChenXOR} and \cite{MaddahAliCentralized} for $M=1/K$ and $M={t^*}N/K$, respectively. In the following corollary, the improved delivery rate-cache capacity trade-off for $1/K \le M \le {t^*}N/K$ is presented.

\begin{corollary} \label{ExtendedImprovementDelRateCol}
The delivery rate-cache capacity trade-off 
\begin{equation} \label{ExtendedImprovementDelRate}
R_{c}(M) = 
\begin{cases}
{ - N\left( {\frac{1}{2}M - 1 + \frac{1}{2K}} \right),\; \quad \qquad \frac{1}{K} \le M \le \frac{N-1}{K}},\\
\frac{1}{\left( {{t^*} - 1} \right)N + 1}\left( {\frac{{K\left( {K - {t^*}} \right)}}{{{t^*} + 1}} - N\left( {K - \frac{N}{2}} \right)} \right)\\
\quad \; \; \left( {M - \frac{{N - 1}}{K}} \right) + N - \frac{{{N^2}}}{{2K}}, \; \frac{{N - 1}}{K} \le M \le \frac{{{t^*}N}}{K}
\end{cases}
\end{equation}
is achievable in a centralized caching system with a database of $N$ files, and $K > N$ users, each equipped with a cache of normalized capacity $M$. 
\end{corollary}

In the following theorem, which is proved in Appendix \ref{ProofMultGap}, we show that the delivery rate-cache capacity trade-off $R_c(M)$ is within a constant multiplicative factor of $2$ of the optimal delivery rate $R^*(M)$ for $1/K \le M \le (N-1)/K$, when $K>N\ge3$.  

\begin{theorem}\label{MultiplicativeGap}
For a caching system with $N$ files, and $K$ users, satisfying $K > N \ge 3$, and a cache capacity of $M \in \left[ 1/K, (N-1)/K \right]$, we have
\begin{equation} \label{MultGapTheorem}
\frac{{{R_c}\left( M \right)}}{{{R^*}\left( M \right)}} \le 2, 
\end{equation}
where $R_c(M)$ is the delivery rate achieved by the proposed coded caching scheme.   
\end{theorem}

For all values of $N$ and $K$, and all cache capacities $0 \le M \le N$, the delivery rate of the centralized coded caching scheme proposed in \cite{MaddahAliCentralized} is shown to be within a constant factor of $12$ and $8$ of the optimal delivery rate by utilizing the lower bounds derived in \cite{MaddahAliCentralized} and \cite{SenguptaCaching}, respectively. Therefore, the proposed caching scheme reduces the best multiplicative gap in the literature by a factor of $4$ for cache capacities satisfying $1/K \le M \le (N-1)/K$, when $K > N \ge 3$, achieved through memory-sharing between the proposed scheme for $M=(N-1)/K$ and the scheme of \cite{ZhiChenXOR} for $M = 1/K$. Note that, the centralized coded caching scheme studied in \cite{ZhiChenXOR} is optimal for cache capacities $M \le 1/K$, when $N \le K$.  

\subsection{Numerical Comparison with the State-of-the-Art}\label{SimSecScheme}
\begin{figure}[!t]
\centering
\includegraphics[scale=0.51]{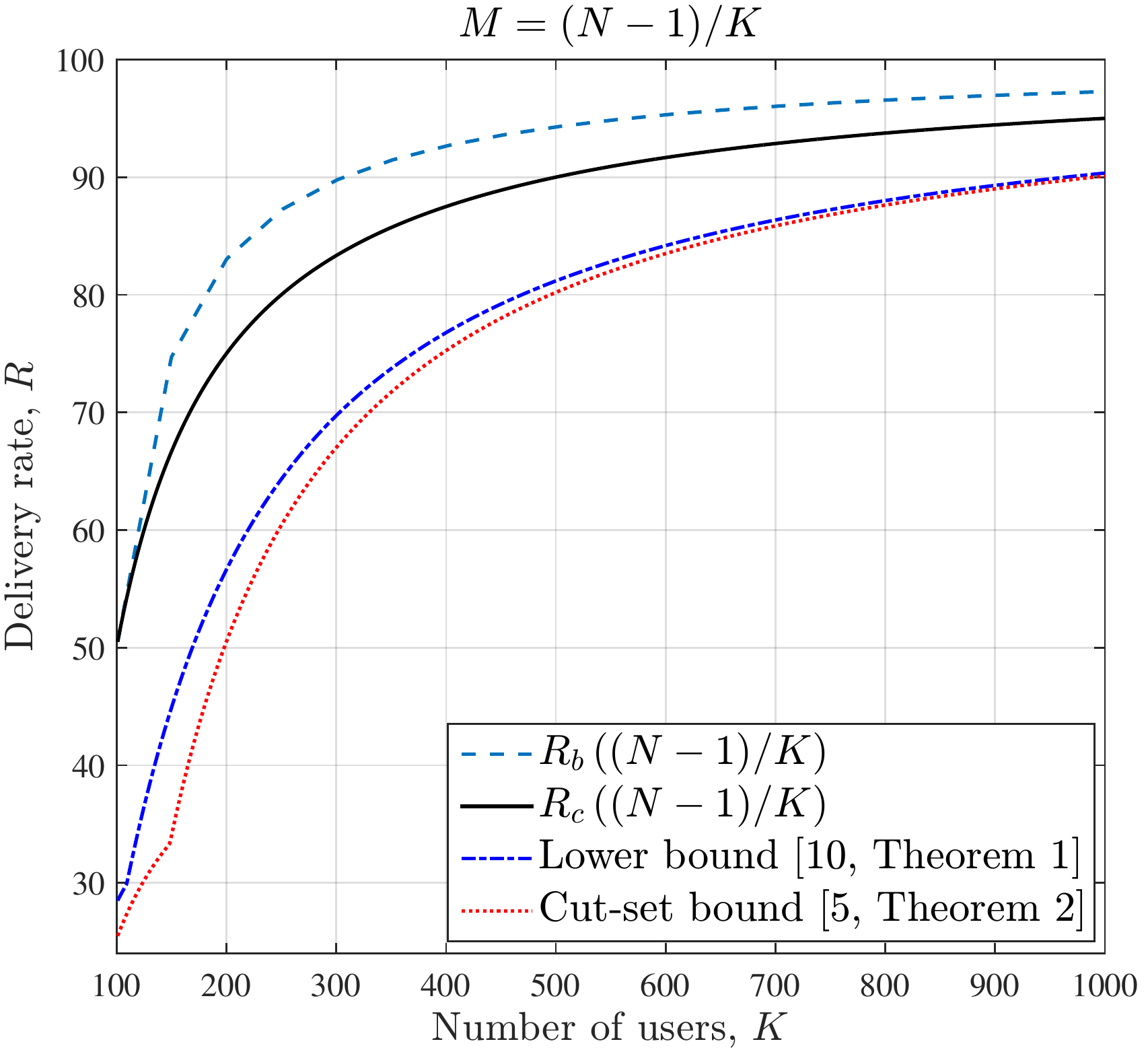}
\caption{Delivery rate for a caching system with $N=100$ files as a function of the number of users, $K \in \left[ 101,1000 \right]$ when $M=(N-1)/K$.} 
\label{N100_K}
\end{figure}

In this section, the delivery rate of the proposed scheme is compared numerically with the state-of-the-art results. In Fig. \ref{N100_K}, the delivery rate of the proposed scheme for $M=(N-1)/K$, i.e., $R_c\left((N-1)/K\right)$ given by \eqref{TheoremSecondSchemeDeliveryRate}, is compared with that of the state-of-the-art for the same cache capacity, i.e., $R_b\left((N-1)/K\right)$ given by \eqref{RbTStar}, as a function of $K \in \left[ 101,1000 \right]$ when $N=100$. It can be seen that for the whole range of $K$ values under consideration, the proposed coded caching scheme outperforms the MNC scheme, and the improvement is more noticeable for relatively moderate values of $K$. We also include in the figure two lower bounds on the delivery rate, the bound derived in \cite[Theorem 1]{SenguptaCaching} and the cut-set based lower bound \cite[Theorem 2]{MaddahAliCentralized}. Despite the improvement, there is still a large gap to the lower bounds. We believe that this gap is largely due to the looseness of the lower bound, but further improvements on the achievable delivery rate may also be possible. We are currently working on reducing this gap in both directions.    

%\begin{figure}[!t]
%\centering
%\includegraphics[scale=0.8]{N100_Kvaries.pdf}
%\caption{Delivery rate for a caching system with $N=100$ files as a function of the number of users, $K \in \left[ 101,1000 \right]$ when $M=(N-1)/K$.} 
%\label{N100_K}
%\end{figure}

\begin{figure}[!t]
\centering
\includegraphics[scale=0.55]{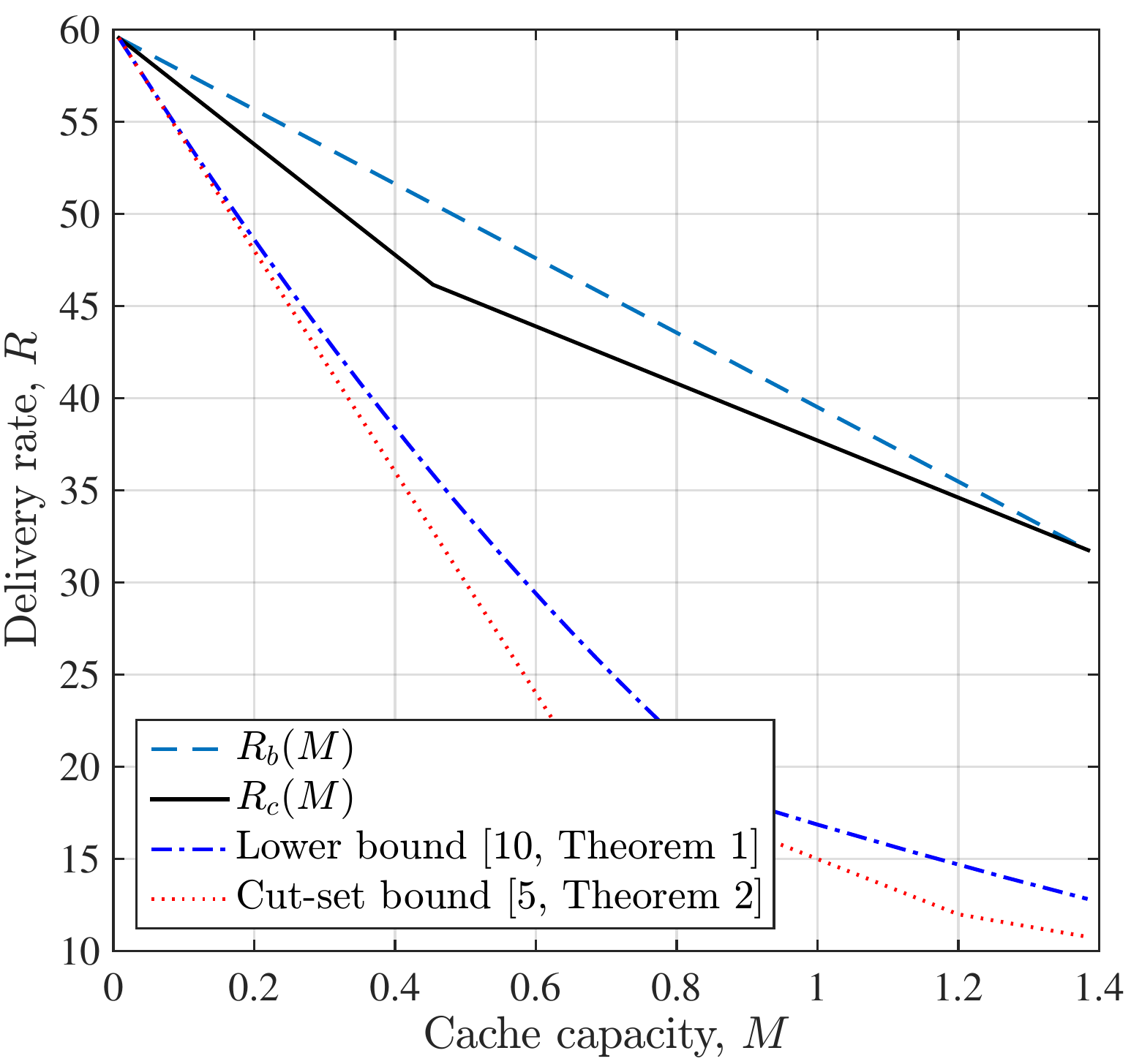}
\caption{Delivery rate-cache capacity trade-off for a caching system with $N=60$ files and $K = 130$ users when $1/K \le M \le t^*N/K$, for which, according to \eqref{TStar}, $t^*=3$.} 
\label{N60_K130}
\end{figure}

In Fig. \ref{N60_K130}, we compare the delivery rate-cache capacity trade-off achieved by the proposed coded caching strategy, $R_c(M)$, with the trade-off acheved by the MNC scheme, $R_b(M)$, when $N=60$ and $K=130$. In the figure, we focus on the cache capacity values $1/K \le M \le t^*N/K$ for which the proposed scheme outperforms the state-of-the-art. Note that, for this setting, based on \eqref{TStar}, we have $t^* = 3$.  Observe that the proposed scheme requires less data to be transmitted by the server over the shared link in the delivery phase for all cache capacity values satisfying $1/K < M < 3N/K$. We also include in the figure the two lower bounds on the delivery rate derived in \cite[Theorem 1]{SenguptaCaching} and \cite[Theorem 2]{MaddahAliCentralized}.

%\begin{figure}[!t]
%\centering
%\includegraphics[scale=0.8]{N60_K130.pdf}
%\caption{Delivery rate-cache capacity trade-off for a caching system with $N=60$ files and $K = 130$ users when $1/K \le M \le t^*N/K$, for which, according to \eqref{TStar}, $t^*=3$.} 
%\label{N60_K130}
%\end{figure}

\section{Conclusions}\label{Conc}
We have proposed a novel centralized coded caching scheme for a content delivery network consisting of a server delivering $N$ popular files, each of length $F$ bits, to $K$ users, each with a cache of capacity $MF$ bits. The proposed coded caching strategy, which is valid for all values of $N$ and $K$, utilizes coded delivery, and creates symmetry between the delivered portions of different files in the delivery phase by partitioning files into different pieces. The delivery phase exploits both coded and uncoded transmission of various pieces of contents, carefully created to retain the symmetry across users and files. The delivery rate achieved by the proposed scheme for a cache capacity of $M=(N-1)/K$ is given by $R_c=N\left( 1-N/(2K) \right)$ for $N < K$, which is shown to be lower than the state-of-the-art obtained by memory-sharing between the coded caching schemes proposed in \cite{MaddahAliCentralized} and \cite{ZhiChenXOR}. We have then extended the improvement to a larger range of cache capacities through memory-sharing between the proposed scheme and the best achievable scheme in the literature to obtain an order-optimal achievable delivery rate, which is shown to be within a constant multiplicative factor of $2$ of the theoretically optimal delivery rate for cache capacities satisfying $1/K \le M \le (N-1)/K$, when $K > N \ge 3$. 

We note that the caching scheme proposed in this paper places coded contents in the user cache memories, similarly to the scheme proposed in \cite{ZhiChenXOR}, and unlike the uncoded cache placement used in \cite{MaddahAliCentralized} and all the other follow-up works in the literature. We observe that, if the number of users in the system is more than the number of files, coded cache placement outperforms uncoded cache placement when the cache capacities are limited. We also note that, in the coded delivery formulation considered here, the total number of bits that need to be delivered over the shared link scales with $F$, the size of each content, which is assumed to be very large in this work. Therefore, the obtained gain in the delivery phase can be significant in terms of the number of delivered bits. 

\appendices

\section{Proof of Theorem \ref{TheoremSecondSchemeDelRate}}\label{ProofSecondTheorem}
To find the delivery rate of the proposed scheme, the delivery rate for each part of the delivery phase is calculated separately. 

Having received the bits sent in the first part of the delivery phase presented in Algorithm \ref{Part1cgeneralSecond}, we would like each user to recover all the subfiles in its cache that have been cached in the XOR-ed form during the placement phase. However, to achieve this, we transmit pieces of the files that are not requested by that user. For example, for user $U_k$ in group $G_i$ with demand $W_i$, $i=1, ..., N$ and $k=S_{i-1}+1, ..., S_i$, we deliver $(N-1)$ different pieces corresponding to $(N-1)$ different files (except file $W_i$) to retrieve all the subfiles $W_{l,k}$, for $l=1, ..., N$. Since there are $K$ users, a total of $K(N-1)$ different pieces, each of length $\frac{F}{K(N-1)}$ bits, are sent over the shared link in the first part of the delivery phase. As a result, the delivery rate of part 1 of the delivery phase is $R_{c_1}=1$.

In part 2 of the proposed delivery phase provided in Algorithm \ref{Part2cgeneralSecond}, for the users in each group $G_i$, $\left( K_i-1 \right)$ XOR-ed contents $\mathop  \bigcup \limits_{k = S_{i - 1} + 1}^{S_{i}-1} \left( {{W_{i,k}} \oplus {W_{i,k + 1}}} \right)$ are transmitted over the shared link, enabling all the users in group $G_i$ to recover the subfiles $W_{i,S_{i-1}+1}, \ldots, W_{i,S_i}$. Hence, a total of $\sum\limits_{i = 1}^N {\left( {{K_i} - 1} \right)}$ XOR-ed contents, each of size $F/K$ bits, are delivered over the shared link, which results in a delivery rate of
\begin{equation}\label{DelRateAnalSecSchemeSecPart}
{R_{{c_2}}} = \frac{1}{K}\sum\limits_{i = 1}^N {\left( {{K_i} - 1} \right)}  = 1 - \frac{N}{K}
\end{equation}
for the second part of the delivery phase.

Finally, Algorithm \ref{Part3cgeneralSecond} corresponds to the last part of the proposed delivery scheme, which enables file exchanges between the users in groups $G_i$ and $G_j$, for $i=1, \ldots, N-1$ and $j=i+1, \ldots, N$. There are $(N-2)$ missing pieces of the file requested by users in group $G_i$ ($G_j$) that are located in the cache of each of the users in group $G_j$ ($G_i$) with indexes ${l_1} \in \left[ {N - 1} \right]\backslash \left\{ {{m_{i,{S_j}}}} \right\}$ (${l_2} \in \left[ {N - 1} \right]\backslash \left\{ {{m_{j,{S_i}}}} \right\}$). Note that, we have $(N-2)$ missing pieces rather than $(N-1)$ as one piece was delivered in part 1 of the delivery scheme. For the piece with index $l_1$ and the piece with index $l_2$, the server delivers $\mathop  \bigcup \limits_{k = S_{j - 1} + 1}^{S_j - 1} \left( {{W_{i,k}^{\left( l_1 \right)}} \oplus {W_{i,k + 1}^{\left( l_1 \right)}}} \right)$, $\mathop  \bigcup \limits_{k = S_{i - 1} + 1}^{S_i - 1} \left( {W_{j,k}^{\left( l_2 \right)}\oplus W_{j,k+1}^{\left( l_2 \right)}} \right)$, and $W_{i, S_j}^{\left( l_1 \right)} \oplus W_{j, S_i}^{\left( l_2 \right)}$, which enables all the users in group $G_i$ to recover the pieces ${W_{i,S_{j-1}+1}^{\left( l_1 \right)}}, ..., {W_{i,S_{j}}^{\left( l_1 \right)}}$, and also all the users in group $G_j$ to recover the pieces $W_{j,S_{i-1}+1}^{\left( l_2 \right)}, ..., W_{j,S_{i}}^{\left( l_2 \right)}$, by delivering a total of $\left( {{K_i} + {K_j} - 1} \right)$ XOR-ed contents, each of size $\frac{F}{K(N-1)}$ bits. As a result, the delivery rate of the third part is given by
\begin{align}\label{DelRateAnalSecSchemeThirdPart}
{R_{{c_3}}} &= \frac{{(N - 2)}}{{K\left( {N - 1} \right)}}\sum\limits_{i = 1}^{N - 1} {\sum\limits_{j = i + 1}^N {\left( {{K_i} + {K_j} - 1} \right)} }  \nonumber\\
&= \left( {N - 2} \right)\left( {1 - \frac{N}{{2K}}} \right). 
\end{align}

By adding up the delivery rate of the three parts, the following delivery rate is achieved:
\begin{equation}\label{DelRateAnalSecSchemeProof}
{R_{{c}}}\left( {\frac{{N - 1}}{K}} \right) = R_{c_1}+R_{c_2}+R_{c_3} = N\left( {1 - \frac{N}{{2K}}} \right),
\end{equation}
which completes the proof of Theorem \ref{TheoremSecondSchemeDelRate}.

\section{Proof of $R_{c}\left((N-1)/K\right) \le R_{b}\left((N-1)/K\right)$}\label{ProofImprovementSecondScheme}
Since both $R_{c}(M)$, given by \eqref{ExtendedImprovementDelRate}, and $R_{b}(M)$, given by \eqref{RbTStar}, starts from $M=1/K$ with the same rate of $N-N/K$ (utilizing the scheme proposed in \cite{ZhiChenXOR}), to show that $R_{c}\left((N-1)/K\right) \le R_{b}\left((N-1)/K\right)$ it suffices to prove that the slope of $R_{c}\left(M\right)$ at $M=1/K$ is not larger than that of $R_{b}\left(M\right)$ at the same point. Observe that the slopes of $R_{c}\left(M\right)$ and $R_{b}\left(M\right)$ at $M=1/K$ are $-N/2$ and $\alpha_{t^*}$, determined according to \eqref{ExtendedImprovementDelRate} and \eqref{RbTStar}, respectively. Due to the difficulty of characterizing $\alpha_{t^*}$ explicitly, we instead show that 
\begin{equation}\label{SlopeDefCom}
{\alpha _t} \ge  - \frac{N}{2}, \quad \forall t \in \left[K \right],
\end{equation}
which concludes that $\alpha_{t^*} \ge -N/2$. We first define a function $g:\left( \left( \mathcal Z^+ , \mathcal Z^+ , \left[ K \right] \right) \to \mathcal R \right)$, where $\mathcal R$ denotes the set of real numbers, as follows:
\begin{equation}\label{DefFuncG}
g\left( {N,K,t} \right) \buildrel \Delta \over = {\alpha _t} + \frac{N}{2},
\end{equation}
and the goal is to illustrate that $g\left( {N,K,t} \right) \ge 0$, which is equivalent to
\begin{equation}\label{ProofImrpovement1}
K\left( {K - t} \right) + N\left( {t + 1} \right)\left( {\frac{{tN}}{2} - K + \frac{1}{2}} \right) \ge 0.
\end{equation}
Inequality \eqref{ProofImrpovement1} can be re-written as follows:
\begin{align}\label{ProofImrpovement2}
&{\left( {K - \frac{1}{2}\left( {t + N\left( {t + 1} \right)} \right)} \right)^2} - \frac{1}{4}{\left( {t + N\left( {t + 1} \right)} \right)^2} \nonumber\\
& \qquad \qquad \qquad \qquad \qquad+ \frac{1}{2}N\left( {t + 1} \right)\left( {tN + 1} \right) \ge 0,
\end{align}
and after some algebraic manipulations, we have $g\left( {N,K,t} \right) \ge 0$, if and only if (iff)
\begin{align}\label{ProofImrpovement3}
h\left( {N,K,t} \right) \buildrel \Delta \over =& {\left( {K - \frac{1}{2}\left( {t + N\left( {t + 1} \right)} \right)} \right)^2} \nonumber\\
& + \frac{1}{4}\left( {{t^2} - 1} \right)N\left( {N - 2} \right) - \frac{1}{4}{t^2} \ge 0.
\end{align}
Observe that for $t=1$, we have 
\begin{equation}\label{ProofImrpovement4}
h\left( {N,K,1} \right) = {\left( {K - N - \frac{1}{2}} \right)^2} - \frac{1}{4} \ge 0.
\end{equation}
Furthermore, for $t \ge 2$ and $N \ge 3$, we have
\newcommand\myfirstinequality{\mathrel{\overset{\makebox[0pt]{\mbox{\normalfont\tiny\sffamily (a)}}}{\ge}}}
\begin{equation}\label{ProofImrpovement5}
h\left( {N,K,t} \right) \ge \frac{1}{4}\left( {{t^2} - 1} \right)N\left( {N - 2} \right) - \frac{1}{4}{t^2} \ge 0,
\end{equation}
Consequently, we have 
\begin{equation}\label{ProofImrpovement6}
g\left( {N,K,t} \right) \ge 0, \quad \mbox{for $N \ge 3$},
\end{equation}
which results in $R_{c}\left((N-1)/K\right) \le R_{b}\left((N-1)/K\right)$ for $N \ge 3$. 

\section{Proof of Theorem \ref{MultiplicativeGap}}\label{ProofMultGap}
We consider two distinct cases, when $N$ is an even number, and when it is an odd number, and prove that for both cases, the multiplicative factor between the proposed achievable delivery rate $R_c(M)$ and $R^*(M)$ is at most $2$ when $K > N \ge 3$, and $1/K \le M \le (N-1)/K$. According to the lower bound on the delivery rate derived in \cite[Theorem 1]{SenguptaCaching}, we have
\begin{align}\label{ProofMultGap0}
&{R^*}\left( M \right) \ge {R_{\rm{LB}}}\left( M \right) \buildrel \Delta \over =\nonumber\\
&\mathop {\max }\limits_{\scriptstyle s \in \left\{ {1,...,K} \right\},\hfill\atop
\scriptstyle l \in \left\{ {1,...,\left\lceil {\frac{N}{s}} \right\rceil } \right\}\hfill} \frac{1}{l}\left\{ {N - sM - \frac{{\mu {{\left( {N - ls} \right)}^ + }}}{{s + \mu }} - {{\left( {N - Kl} \right)}^ + }} \right\},
\end{align}
where $\mu \buildrel \Delta \over = \min \left\{ {\left\lceil {\left( {N - ls} \right)/l} \right\rceil ,K - s} \right\}$, and ${\left( x \right)^ + } \buildrel \Delta \over = \max \left\{ {x,0} \right\}$. 

First, assume that $N$ is an even number. By setting $s=N/2$ and $l=1$ in \eqref{ProofMultGap0}, for $K>N$, we have
\begin{equation}\label{ProofMultGap1}
{R^*}\left( M \right) \ge N - \frac{N}{2}M - \frac{{\mu N}}{{N + 2\mu }},
\end{equation}
where
\begin{equation}\label{ProofMultGap2}
\mu  = \min \left\{ {\left\lceil {\frac{N}{2}} \right\rceil ,K - \frac{N}{2}} \right\} = \frac{N}{2},
\end{equation}
which follows since $K > N$ and $N$ is even. By substituting $\mu$ from \eqref{ProofMultGap2} to \eqref{ProofMultGap1}, we have
\begin{equation}\label{ProofMultGap3}
{R^*}\left( M \right) \ge \frac{N}{4}\left( {3 - 2M} \right).
\end{equation}
According to \eqref{ExtendedImprovementDelRate}, for $1/K \le M \le (N-1)/K$, we have
\begin{equation}\label{ProofMultGap4}
\frac{{{R_c}\left( M \right)}}{{{R^*}\left( M \right)}} \le \frac{{2\left( {2 - M - 1/K} \right)}}{{3 - 2M}}.
\end{equation}
Note that, the expression on the right hand side of inequality \eqref{ProofMultGap4} is an increasing function of $M$ for $K \ge 2$. Setting the cache capacity to its maximum value, $(N-1)/K$, we have   
\begin{equation}\label{ProofMultGap5}
\frac{{{R_c}\left( M \right)}}{{{R^*}\left( M \right)}} \le \frac{{2\left( {2K - N} \right)}}{{3K - 2N + 2}} \le \frac{{2\left( {2K - N} \right)}}{{3K - 2N }}.
\end{equation}
The last expression above is a decreasing function of $K$, so by letting $K=N+1$, we have
\begin{equation}\label{ProofMultGap6}
\frac{{{R_c}\left( M \right)}}{{{R^*}\left( M \right)}} \le \frac{{2\left( {N+2} \right)}}{{N+3}} \le 2.
\end{equation}
Now, consider $N$ is an odd number. For $K>N$, we can set $s=(N-1)/2$ and $l=1$ in the lower bound of \eqref{ProofMultGap0} to find
\begin{equation}\label{ProofMultGap7}
{R^*}\left( M \right) \ge N - \frac{{N - 1}}{2}M - \frac{{\left( {N + 1} \right)\mu }}{{N - 1 + 2\mu }},
\end{equation}
where
\newcommand\mysecondequality{\mathrel{\overset{\makebox[0pt]{\mbox{\normalfont\tiny\sffamily (a)}}}{=}}}
\begin{equation}\label{ProofMultGap8}
\mu  = \min \left\{ {\left\lceil {N - \frac{{N - 1}}{2}} \right\rceil ,K - \frac{{N - 1}}{2}} \right\} = \frac{{N + 1}}{2}.
\end{equation}
Thus, we have  
\begin{equation}\label{ProofMultGap9}
{R^*}\left( M \right) \ge N - \frac{{N - 1}}{2}M - \frac{{{{\left( {N + 1} \right)}^2}}}{{4N}}.
\end{equation}
For $1/K \le M \le (N-1)/K$, we can obtain
\begin{equation}\label{ProofMultGap10}
\frac{{{R_c}\left( M \right)}}{{{R^*}\left( M \right)}} \le \frac{{2 - M - 1/K}}{{2 - \left( {1 - \frac{1}{N}} \right)M - \frac{1}{2}{{\left( {1 + \frac{1}{N}} \right)}^2}}}.
\end{equation}
In the following, we show that the function $f:\left( \left( \left[ 1/K,(N-1)/K \right] \right) \to \mathcal R \right)$, defined as
\begin{equation}\label{ProofMultGap11}
f(M) \buildrel \Delta \over = \frac{{2 - M - 1/K}}{{2 - \left( {1 - \frac{1}{N}} \right)M - \frac{1}{2}{{\left( {1 + \frac{1}{N}} \right)}^2}}},
\end{equation}
is an increasing function of $M$ for $K>N \ge 3$. We have
\begin{equation}\label{ProofMultGap12}
\frac{{df}}{{dM}} = \frac{{\left( {1 - \frac{1}{N}} \right)\left( {\frac{{\left( {N - 1} \right)K - 2N}}{{2NK}}} \right)}}{{{{\left( {2 - \left( {1 - \frac{1}{N}} \right)M - \frac{1}{2}{{\left( {1 + \frac{1}{N}} \right)}^2}} \right)}^2}}} \ge 0,
\end{equation}
where the last inequality in \eqref{ProofMultGap12} holds for $K>N\ge 3$. Hence, we have
\begin{align}\label{ProofMultGap13}
\frac{{{R_c}\left( M \right)}}{{{R^*}\left( M \right)}} \le& f\left( {\frac{{N - 1}}{K}} \right) \nonumber\\
& \quad = \frac{{2 - N/K}}{{2 - \left( {1 - \frac{1}{N}} \right)\left( {\frac{{N - 1}}{K}} \right) - \frac{1}{2}{{\left( {1 + \frac{1}{N}} \right)}^2}}}.
\end{align}
Now the goal is to prove that $f \left( (N-1)/K \right) \le 2$, for $K>N\ge 3$, when $N$ is an odd number. After some simplification, it can be seen that $f \left( (N-1)/K \right) \le 2$ iff
\begin{equation}\label{ProofMultGap14}
p\left( {N,K} \right) \buildrel \Delta \over = 1 - \frac{{N - 4}}{K} - \frac{2}{{KN}} - \frac{1}{{{N^2}}} - \frac{2}{N} \ge 0.
\end{equation}
Observe that, for $N \ge 4$, $p(N,K)$ is an increasing function of $K$. Thus, replacing $K=N+1$, we have
\begin{equation}\label{ProofMultGap15}
p\left( N,K \right) \ge \frac{{N\left( {3N - 5} \right) - 1}}{{{N^2}\left( {N + 1} \right)}} \ge 0,
\end{equation}
where the above inequality follows since $N \ge 4$. To complete the proof we need to show that $p\left( 3,K \right) \ge 0$, for $K \ge 4$. We have 
\begin{equation}\label{ProofMultGap16}
p\left( 3,K \right) = \frac{2}{9} + \frac{1}{{3K}} \ge 0,
\end{equation}
which completes the proof for odd $N$ values. As a result, for $1/K \le M \le (N-1)/K$, the multiplicative gap between the delivery rate of the proposed scheme and the optimal delivery rate is at most $2$ for all $K>N\ge 3$.

\bibliographystyle{IEEEtran}
%\addbibresource{ReportTCOMM1.bib}
\bibliography{Report}

\begin{IEEEbiography}[{\includegraphics[width=1in,height=1.25in,clip,keepaspectratio]{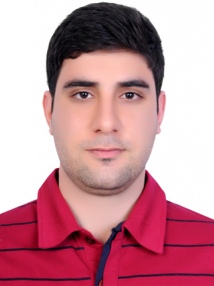}}]{Mohammad Mohammadi Amiri}

received the B.Sc. degree in Communication Systems from the Iran University of Science and Technology in 2011 and the M.Sc. degree in Communication Systems engineering from the University of Tehran in 2014 both with highest rank in classes. Currently, he is a research postgraduate at Imperial College London under the supervision of Dr. G\"und\"uz. His research interests include information and coding theory, wireless communications, MIMO systems, cooperative networks, cognitive radio, and signal processing.

\end{IEEEbiography}

\begin{IEEEbiography}[{\includegraphics[width=1in,height=1.25in,clip,keepaspectratio]{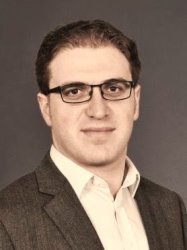}}]{Deniz G\"und\"uz}

[S'03-M'08-SM'13] received the B.S. degree in electrical and electronics engineering from METU, Turkey in 2002, and the M.S. and Ph.D. degrees in electrical engineering from NYU Polytechnic School of Engineering in 2004 and 2007, respectively. After his PhD, he served as a postdoctoral research associate at Princeton University, and as a consulting assistant professor at Stanford University. He was a research associate at CTTC in Barcelona, Spain until September 2012, when he joined the Electrical and Electronic Engineering Department of Imperial College London, UK, as a Lecturer. Currently he is a Reader in the same department.

Dr. G\"und\"uz is an Editor of the IEEE Transactions on Communications, and IEEE Transactions on Green Communications and Networking. He is the recipient of a Starting Grant of the European Research Council (ERC), the 2014 IEEE Communications Society Best Young Researcher Award for the Europe, Middle East, and Africa Region, Best Paper Award at the 2016 IEEE Wireless Communications and Networking Conference (WCNC), and the Best Student Paper Award at the 2007 IEEE International Symposium on Information Theory (ISIT). He served as the General Co-chair of the 2016 IEEE Information Theory Workshop, and the 2012 IEEE European School of Information Theory (ESIT). His research interests lie in the areas of communication theory and information theory with special emphasis on joint source-channel coding, multi-user networks, energy efficient communications and privacy in cyber-physical systems.

\end{IEEEbiography}

\end{document}